\documentclass[nofootinbib,notitlepage,superscriptaddress,aps,prd]{revtex4-1}

\usepackage[utf8x]{inputenc}
\usepackage{amsmath}
\usepackage{amssymb}
\usepackage{amsfonts}
\usepackage{graphicx}
\usepackage{bbm}
\usepackage{bm}

\begin{document}

\title{Multi-particle systems in $\kappa$-Poincaré inspired by 2+1D gravity}

\author{Jerzy Kowalski-Glikman}
\email{jerzy.kowalski-glikman@ift.uni.wroc.pl}\affiliation{Institute
for Theoretical Physics, University of Wroc\l{}aw, Pl.\ Maksa Borna
9, Pl--50-204 Wroc\l{}aw, Poland}

\author{Giacomo Rosati}
\email{giacomo.rosati@ift.uni.wroc.pl}\affiliation{Institute
for Theoretical Physics, University of Wroc\l{}aw, Pl.\ Maksa Borna
9, Pl--50-204 Wroc\l{}aw, Poland}

\begin{abstract}
Inspired by a Chern-Simons description of 2+1D gravity coupled to
point particles  we propose a new Lagrangian
of a multiparticle system living in $\kappa$-Minkowski/$\kappa$-Poincaré
spacetime. We derive the dynamics of interacting
particles with $\kappa$-momentum space, alternative to the one proposed
in the ``principle of relative locality'' literature.
The model that we obtain takes into account of the nonlocal topological interactions between the particles, so that the effective multi-particle action is not a sum of their free actions.
In this construction the locality of particle processes is naturally
implemented, even for distant observers. In particular a particle
process is characterized by a local deformed energy-momentum conservation
law. The spacetime transformations are generated by total charges/generators for the composite particle system, and leave unaffected the locality of individual particle processes.
\end{abstract}

\maketitle

\section{Introduction}

Gravity in 2+1 dimensions \cite{Witten:1988hc}, \cite{Achucarro:1987vz}
is a remarkable theory. It is described by a topological field theory
and therefore it does not possess any dynamical degrees of freedom,
so that the gravitational waves and Newtonian interactions of particles
are not present. In spite of its apparent dullness, this theory contains
rich physics when coupled to point particles or fields. The reason
is the very simplicity of 2+1D gravity. In the case of the gravitating
particle(s), one can find explicitly the form of the gravitational
field and then substitute it back to the action to obtain the effective
particle action that includes exactly the gravitational back-reaction
\cite{deSousaGerbert:1990yp}, \cite{Matschull:1997du}, \cite{Meusburger:2003ta}.
In the case of the quantum theory of a field coupled to gravity, in
the path-integral formalism one can exactly integrate out all the
gravitational degrees of freedom, obtaining an effective field theory
\cite{Freidel:2005bb}, \cite{Freidel:2005me}. In both cases the
resulting, effective theories share a couple of important properties,
their relativistic symmetries are deformed, spacetimes are non-commutative,
and momentum spaces possess a nontrivial geometry. All these effects
are characterized by a single mass scale $\kappa$, which is inverse
proportional to the 2+1 dimensional Newton's constant, and disappears
when the gravitational coupling goes to zero. The above are the features
of a class of theories belonging to the general relative locality
framework~\cite{AmelinoCamelia:2011bm} and therefore gravity in
2+1 dimensions coupled to particles and/or fields serves as a basic
explicit example of this class of theories (cf.~\cite{AmelinoCamelia:2012rz} for an explicit example).

It has been shown~\cite{AmelinoCamelia:2010qv,AmelinoCamelia:2011cv}
that in a relativistic
theory (in the sense of DSR~\cite{AmelinoCamelia:2000mn,AmelinoCamelia:2000ge,KowalskiGlikman:2001gp}),
the introduction of the invariant scale $\kappa$, with dimension
of momentum, characterizing the momentum space geometry, makes it
necessary to relax the notion of locality, which becomes a relative,
observer-dependent notion. In this respect a particle process is described
as local only in the coordinates of an observer sitting ``at'' the
process. A distant observer, in her coordinatization, describes the
process as ``non-local''. Technically, these relative locality effects
are encoded in the non-linearity of the relativistic transformation
laws connecting the coordinates of inertial observers.

A general feature of theories with (DSR) deformed relativistic symmetries, like are relative locality theories, is that a deformed composition rule of
particles momenta is needed to implement energy/momentum conservation
in a particle process.  For instance, for a process involving two
incoming particles with momenta $p_{(1)},p_{(2)}$, and two outgoing
particles with momenta $p_{(3)},p_{(4)}$, we can interpret the composite
system of particles characterized by the total momentum $p^{(in)tot}=p_{(1)}\oplus p_{(2)}$,
as the in-state of a particle process, and the composite system of
particles characterized by the total momentum $p^{(out)tot}=p_{(3)}\oplus p_{(4)}$,
as the outgoing state, where the $\oplus$ encodes the deformed summation
rule. The conservation of energy/momentum of the process is then implemented
by the constraint
\begin{equation}
p^{(in)tot}=p^{(out)tot}.
\end{equation}
In~\cite{KowalskiGlikman:2004tz,AmelinoCamelia:2011bm} the deformed summation
law is associated to the nontrivial geometry of momentum space.
This deformed momentum summation law characterizing particle processes
 is implemented through a suitable
boundary term, strictly related to the translational symmetries of
the theory, which was a subject of the studies presented in~\cite{AmelinoCamelia:2011nt,Carmona:2011wc}.
It was shown that the requirement for the theory to be relativistic
restricts the class of possible boundary terms. One of the outcomes
of these studies was that the relative locality emerges from the description
of the spacetime position of the event of interaction.

The basic property of the models investigated in the framework of
relative locality (see~\cite{Kowalski-Glikman:2013rxa} for a recent
review) was that, apart from the vertex, the particles' kinematics is
governed by a free, albeit in general deformed, action. Such construction
is based on the intuition stemming from a construction based on Feynman-like diagrams,
where we have to do with free particles on the diagram lines interacting
at a final number of vertices. However, the Feynman-like diagrams are aimed
to describe local processes and it is not clear if they are really
the best point of departure in the case nonlocal theories. In particular,
it is well established in the context of the theory of particles coupled
to 2+1 gravity that, after solving out the gravity degrees of freedom,
the effective multi-particles action is \underline{not} a sum of
their free actions, and that one has to do with a nontrivial, nonlocal topological
interactions between the particles~\cite{Meusburger:2003ta}.

In this paper we generalize this construction to the case of a two-particle
system in 3+1 spacetime dimensions, and choose as illustrative example
the case in which the momentum space of the particle is the $AN(3)$
group i.e., (a part of) de Sitter space. It is well known \cite{KowalskiGlikman:2003we,KowalskiGlikman:2004tz,Arzano:2010kz}
that in the free particle case this example corresponds to the $\kappa$-Poincaré
deformation \cite{Lukierski:1991pn,Lukierski:1992dt,Majid:1994cy}, which plays an important role in the quantum gravity phenomenology~\cite{AmelinoCamelia:2008qg}.
Here we will find that in the multiparticle case the symmetries of
the system are described by the $\kappa$-Poincaré Hopf algebra. Our
construction differs from the models considered previously~\cite{AmelinoCamelia:2011bm,Gubitosi:2013rna,Carmona:2011wc,AmelinoCamelia:2011nt}.
The system we obtain has the property that under the action of total
momentum the two particle coordinates translate ``rigidly'', by
the same amount. Noting that our model implements translational invariance
in a way compatible with the description of boundary terms achieved
in~\cite{AmelinoCamelia:2011nt,Carmona:2011wc}, we propose a new
expression for the action of two-particle processes, which combines
our braided two-particle Lagrangian and the expression for the boundary
terms in a natural way.

The main novelty of our model lies in the fact that, differently from
the previous formulations, the total momentum of the particle system
is at any time equal to the deformed sum of their individual momenta
and that the locality of particle processes is preserved under translations.
We show these results explicitly for the case of a single vertex, while we postpone a detailed analysis of multiple processes to a forthcoming paper. As it will become clear in the following, we expect relative locality effects to be still present when considering several processes.
We conclude our analysis of spacetime symmetries by showing that one can construct total boost and rotation generators for the composite particle system consistently with its total momentum characterization. As for translations, the boosts and rotations so generated also preserve the locality of a particle process.

\section{2-particle Lagrangian}

Using the Chern-Simons description of 2+1D gravity coupled to point
particles \cite{Meusburger:2003ta}, \cite{Meusburger:2005mg} one
can derive the following expression for the kinetic term of an effective
Lagrangian description of the particle dynamics for a system of two
particles:
\begin{equation}
{\cal L}_{1\oplus2}^{kin}=\left\langle \dot{{\cal P}}_{(1)}{\cal P}_{(1)}^{-1},x_{(1)}\right\rangle +\left\langle \dot{{\cal P}}_{(2)}{\cal P}_{(2)}^{-1},x_{(2)}\right\rangle +\left\langle {\cal P}_{(1)}\dot{{\cal P}}_{(2)}{\cal P}_{(2)}^{-1}{\cal P}_{(1)}^{-1}-\dot{{\cal P}}_{(2)}{\cal P}_{(2)}^{-1},x_{(1)}\right\rangle .\label{eq:kineticTerm}
\end{equation}
In this formula ${\cal P}_{(a)}\in\mathfrak{G}$ is an appropriate
group element corresponding to the (group valued) momentum of the
$a$'th particle, which is equal to the holonomy of the (flat) connection,
representing the gravitational field, around the particle's worldline.
The dual phase space coordinates $x_{(a)}\in\mathfrak{g}^{*}$, belonging
to the vector space dual to the Lie algebra of the group, have a natural
interpretation of the particle positions, and the brackets $\left\langle \cdot\right\rangle $
stand for the gauge invariant inner product. Finally $\tau$ plays
a role of the global external time.
In the framework of 2+1 gravity
the models of this kind have been specifically constructed for momentum
group manifold being either $\mathfrak{G}=SO(2,1)$ group (Anti de
Sitter momentum space \cite{Meusburger:2003ta}, \cite{Meusburger:2005mg}
or $\mathfrak{G}=AN(2)$ group (momentum space equal to the half of
de Sitter space covered by flat cosmological coordinates) \cite{Kowalski-Glikman:2014paa}.

The expression (\ref{eq:kineticTerm}) was, strictly speaking, derived from gravity
only in 2+1 dimensions, but it is well defined in any number of spacetime
dimensions, equal to the dimension of the group in question.
Moreover the Lagrangian (\ref{eq:kineticTerm}) is well defined for any group $\mathfrak{G}$.
Since we want the group manifold be the momentum space, we could, in principle, consider any Lie group of dimension 4.
Here, we take this Lagrangian as a starting point for our investigations in 3+1
dimensions, taking as a group $\mathfrak{G}$ the group $AN(3)$%
\footnote{Since the group $AN(D)$ is defined for arbitrary $D$ and has the
dimension $D+1$, the results of our investigations can be readily
applied to an arbitrary spacetime dimension.%
}. As it is well known~\cite{KowalskiGlikman:2004tz,Arzano:2010kz} this group serves
as the basic building block for theories characterized by $\kappa$-Poincaré
symmetries~\cite{Lukierski:1991pn,Lukierski:1992dt,Majid:1994cy}.

The Lagrangian (\ref{eq:kineticTerm}) is clearly not symmetric under particle swap. This is a general feature of the 2+1 dimensional models (see e.g. \cite{Lo:1993hp}) that we adapt here.  This feature seems rather disturbing at the first sight, but it should be recalled that when the momentum space starts having a nontrivial geometry, the abelian, associative structure of the momentum space of special relativity is lost and a choice of the particle's ordering is necessary. In the earlier works \cite{AmelinoCamelia:2011bm,Gubitosi:2013rna,Carmona:2011wc,AmelinoCamelia:2011nt}  this nontrivial aspect of the theory has been fully hidden in the structure of the interaction vertex; here we implement it already at the level of the free particles wordlines, which, as we will see below, makes it possible to cast the conservation laws into much clearer and less ambiguous form.

We introduce momentum space coordinates $p_{(a)\mu}=p_{(a)\mu}\left(\tau\right)$
by writing the group valued momentum in the form
\begin{equation}
{\cal P}_{(a)}=e^{{\cal X}^{j}p_{(a)j}}e^{{\cal X}^{0}p_{(a)0}},\label{eq:TTR}
\end{equation}
where ${\cal X}^{\mu}$ satisfy the $an(3)$ ($\kappa$-Minkowski
\cite{Lukierski:1993wx,Majid:1994cy}) algebra commutation relations
\begin{equation}
\left[{\cal X}^{i},{\cal X}^{0}\right]=\frac{1}{\kappa}{\cal X}^{i},\qquad\left[{\cal X}^{j},{\cal X}^{k}\right]=0.\label{eq:k-mink}
\end{equation}
Expanding the spacetime coordinates as $x_{(a)}\left(\tau\right)={\cal A}_{\mu}x_{(a)}^{\mu}\left(\tau\right)$,
where ${\cal A}_{\mu}\in an^{*}(3)$ are the generators of the dual
algebra, the inner product is defined by the pairing $\left\langle {\cal X}^{\mu},{\cal A}_{\nu}\right\rangle =\delta_{\nu}^{\mu}$.

Using (\ref{eq:k-mink}) we get useful relations
\begin{equation}
\left[e^{{\cal X}^{j}p_{(a)j}},{\cal X}^{0}\right]=\frac{1}{\kappa}p_{(a)j}{\cal X}^{j}e^{{\cal X}^{j}p_{(a)j}},\qquad\left[{\cal X}^{j},e^{{\cal X}^{0}p_{(a)0}}\right]=\left(e^{p_{(a)0}/\kappa}-1\right)e^{{\cal X}^{0}p_{(a)0}}{\cal X}^{j},\label{eq:comm0-1}
\end{equation}
from which it follows that ($\dot{f}=df/d\tau$)
\begin{equation}
\dot{{\cal P}}_{(a)}=\left({\cal X}^{\mu}\dot{p}_{(a)\mu}+\frac{1}{\kappa}{\cal X}^{j}p_{(a)j}\dot{p}_{(a)0}\right){\cal P}_{(a)}.\label{eq:dtP}
\end{equation}
By using Eq. (\ref{eq:dtP}) and (\ref{eq:comm0-1}), and solving
the inner product, the kinetic term becomes
\begin{equation}
\begin{split} & {\cal L}_{1\oplus2}^{kin}=\left(x_{(1)}^{0}+\frac{1}{\kappa}x_{(1)}^{j}p_{(1)j}\right)\dot{p}_{(1)0}+x_{(1)}^{j}\dot{p}_{(1)j}\\
 & \ \ \ +\left(x_{(2)}^{0}+\frac{1}{\kappa}x_{(2)}^{j}p_{(2)j}-\frac{1}{\kappa}x_{(1)}^{j}\left(\left(1-e^{-\frac{1}{\kappa}p_{(1)0}}\right)p_{(2)j}-p_{(1)j}\right)\right)\dot{p}_{(2)0}+\left(x_{(2)}^{j}-x_{(1)}^{j}\left(1-e^{-\frac{1}{\kappa}p_{(1)0}}\right)\right)\dot{p}_{(2)j}.
\end{split}
\label{eq:kin}
\end{equation}

To complete the construction of the Lagrangian for the system of two
particles we have to impose the mass-shell constraints on each particle.
The mass-shell relation is given by $\kappa$-Poincaré mass Casimir%
\footnote{Notice that this can be associated with the geodesic length $\mu$
in deSitter-momentum-space with the cosmological constant $\kappa^2$ by setting $m_{(a)}^{2}=2\kappa^2\left(\cosh\left(\mu_{(a)}/\kappa\right)-1\right)$.%
}
\begin{equation}
{\cal C}_{(a)}=4\kappa^{2}\sinh\left(\frac{p_{(a)0}}{2\kappa}\right)^{2}-e^{p_{(a)0}/\kappa}\mathbf{p}_{(a)}^{2}.\label{eq:casimir}
\end{equation}
Thus the two-particle Lagrangian is
\begin{equation}
{\cal L}_{1\oplus2}={\cal L}_{1\oplus2}^{kin}+\lambda_{(1)}\left({\cal C}_{(1)}-m_{(1)}^{2}\right)+\lambda_{(2)}\left({\cal C}_{(2)}-m_{(2)}^{2}\right).\label{eq:lagr}
\end{equation}

The equations of motion following from the variation of positions
$x_{(a)}^{\mu}$ are the usual momentum conservation conditions
\begin{equation}
\dot{p}_{(1)\mu}=\dot{p}_{(2)\mu}=0.\label{eq:ConservationMomenta}
\end{equation}
The equations resulting from the variations of momenta $p_{(a)\mu}$
are then
\begin{equation}
\begin{gathered}\dot{x}_{(1)}^{0}=\lambda_{(1)}\left(\frac{\partial{\cal C}_{(1)}}{\partial p_{(1)0}}-\frac{1}{\kappa}\mathbf{p}_{(1)}\cdot\frac{\partial{\cal C}_{(1)}}{\partial\mathbf{p}_{(1)}}\right),\qquad\dot{x}_{(1)}^{j}=\lambda_{(1)}\frac{\partial{\cal C}_{(1)}}{\partial p_{(1)j}},\\
\dot{x}_{(2)}^{0}=\lambda_{(2)}\left(\frac{\partial{\cal C}_{(2)}}{\partial p_{(2)0}}-\frac{1}{\kappa}\mathbf{p}_{(2)}\cdot\frac{\partial{\cal C}_{(1)}}{\partial\mathbf{p}_{(1)}}\right)-\lambda_{(1)}\frac{1}{\kappa}\mathbf{p}_{(1)}\cdot\frac{\partial{\cal C}_{(1)}}{\partial\mathbf{p}_{(1)}},\qquad\dot{x}_{(2)}^{j}=\lambda_{(2)}\frac{\partial{\cal C}_{(2)}}{\partial p_{(2)j}}+\lambda_{(1)}\left(1-e^{-p_{(1)0}/\kappa}\right)\frac{\partial{\cal C}_{(1)}}{\partial p_{(1)j}},
\end{gathered}
\end{equation}
which take the explicit form
\begin{equation}
\begin{gathered}\dot{x}_{(1)}^{0}=2\lambda_{(1)}\left(\kappa\sinh\left(\frac{p_{(1)0}}{\kappa}\right)+\frac{1}{2\kappa}e^{p_{(1)0}/\kappa}\mathbf{p}_{(1)}^{2}\right),\qquad\dot{x}_{(1)}^{j}=-2\lambda_{(1)}e^{p_{(1)0}/\kappa}p_{(1)j},\\
\dot{x}_{(2)}^{0}=2\lambda_{(2)}\left(\kappa\sinh\left(\frac{p_{(2)0}}{\kappa}\right)+\frac{1}{2\kappa}e^{p_{(2)0}/\kappa}\mathbf{p}_{(2)}^{2}\right)+2\lambda_{(1)}\frac{1}{\kappa}e^{p_{(1)0}/\kappa}\mathbf{p}_{(1)}^{2},\\
\dot{x}_{(2)}^{j}=-2\lambda_{(2)}e^{p_{(2)0}/\kappa}p_{(2)j}-2\lambda_{(1)}\left(e^{p_{(1)0}/\kappa}-1\right)p_{(1)j}.
\end{gathered}
\label{eq:EqMotion}
\end{equation}
These equations express the (deformed) relations between velocities
and momenta.

\section{Symplectic structure\label{sec:symplectic}}

One can derive the Poisson brackets by defining the variables canonically
conjugate to the $p_{(a)\mu}$, $\tilde{x}_{(a)}^{\mu}=\partial{\cal L}/\partial\dot{p}_{(a)\mu}$,
such that $\left\{ p_{(a)\mu},\tilde{x}_{(a)}^{\nu}\right\} =\delta_{\mu}^{\nu}$,
$\left\{ \tilde{x}_{(a)}^{\mu},\tilde{x}_{(a)}^{\nu}\right\} =0$.
From Eq. (\ref{eq:kin}), the relation between the $x_{(a)}^{\mu}$
and the $\tilde{x}_{(a)}^{\mu}$ are
\begin{equation}
\begin{gathered}x_{(1)}^{0}=\tilde{x}_{(1)}^{0}-\frac{1}{\kappa}\mathbf{p}_{(1)}\cdot\tilde{\mathbf{x}}_{(1)},\qquad x_{(1)}^{j}=\tilde{x}_{(1)}^{j},\\
x_{(2)}^{0}=\tilde{x}_{(2)}^{0}-\frac{1}{\kappa}\left(\mathbf{p}_{(2)}\cdot\tilde{\mathbf{x}}_{(2)}+\mathbf{p}_{(1)}\cdot\tilde{\mathbf{x}}_{(1)}\right),\qquad x_{(2)}^{j}=\tilde{x}_{(2)}^{j}+\left(1-e^{-p_{(1)0}/\kappa}\right)\tilde{x}_{(1)}^{j},
\end{gathered}
\label{eq:xRepCan}
\end{equation}
The Poisson brackets then are
\begin{equation}
\left\{ p_{(a)0},x_{(a)}^{0}\right\} =1,\quad\left\{ p_{(a)0},x_{(a)}^{j}\right\} =0,\quad\left\{ p_{(a)j},x_{(a)}^{0}\right\} =-\frac{1}{\kappa}p_{(a)j},\quad\left\{ p_{(a)j},x_{(a)}^{k}\right\} =\delta_{j}^{k},\label{eq:PB}
\end{equation}
\begin{equation}
\left\{ p_{(1)0},x_{(2)}^{\nu}\right\} =0,\quad\left\{ p_{(1)j},x_{(2)}^{0}\right\} =-\frac{1}{\kappa}p_{(1)j},\quad\left\{ p_{(1)j},x_{(2)}^{k}\right\} =\delta_{j}^{k}\left(1-e^{-p_{(1)0}/\kappa}\right),\quad\left\{ p_{(2)\mu},x_{(1)}^{\nu}\right\} =0,\label{eq:PBmixed}
\end{equation}
\begin{equation}
\begin{gathered}\left\{ x_{(1)}^{0},x_{(1)}^{j}\right\} =\left\{ x_{(1)}^{0},x_{(2)}^{j}\right\} =\left\{ x_{(2)}^{0},x_{(1)}^{j}\right\} =-\frac{1}{\kappa}x_{(1)}^{j},\quad\left\{ x_{(2)}^{0},x_{(2)}^{j}\right\} =-\frac{1}{\kappa}x_{(2)}^{j},\qquad\left\{ x_{(a)}^{j},x_{(a)}^{k}\right\} =0.\end{gathered}
\label{eq:PBcoord}
\end{equation}
Notice that as a consequence of the Lagrangian (\ref{eq:kin}), the
two-particle phase space gets mixed.

By Eq. (\ref{eq:lagr}), the Hamiltonian is
\begin{equation}
{\cal H}={\cal L}-\sum_{i}\tilde{x}^{\mu}\dot{p}_{\mu}=\sum_{i}\lambda_{(a)}\left({\cal C}_{(a)}-m_{(a)}^{2}\right).
\label{eq:Hamiltonian}
\end{equation}
This Hamiltonian generates the evolution in terms of time $\tau$.
Indeed, using the Poisson brackets defined in this section one can
verify that Eqs. (\ref{eq:ConservationMomenta}) and (\ref{eq:EqMotion})
can be rewritten as
\begin{equation}
\dot{p}_{(a)\mu}=\left\{ {\cal H},p_{\mu}\right\} =0,\qquad\dot{x}_{(a)}^{\mu}=\left\{ {\cal H},x^{\mu}\right\} .
\end{equation}

\section{Rigid translations, the total momentum of the system\label{sec:Rigid-translations}}

The fact that the momenta of individual particles are conserved, Eq.~(\ref{eq:ConservationMomenta}),
implies that also every (even non-linear) combination of them is
conserved. Then, the requirement for the total momentum to be conserved
does not single out a unique generator for translations. In principle,
the conserved charges, and hence the generators of symmetries, associated
with translations, can be chosen arbitrarily. There must be therefore
some other property, besides conservation, constraining the form of
the translations generators. We will show now that the requirement
for the total momentum to generate ``rigid translations'' for the
2-particle system, i.e. to be such that the coordinates of both particle
translate by the same amount, suffices to single out an expression
for the total momentum. 
It is not guaranteed by any means that such total momentum exists in general; for example it does not in the original relative locality model. It is a remarkable property of the model we are considering here, directly related to the non-symmetric form of the Lagrangian (\ref{eq:kineticTerm}), that the total momentum generating rigid transformations does exist and, moreover is exactly the total momentum defined by the $AN(3)$ group product (see eq.\ (\ref{28})).

To see how this argument works let us revisit the Lagrangian\ (\ref{eq:kin})
and express it in terms of the relative position of the particles
$x_{(-)}^{\mu}=1/2\left(x_{(2)}^{\mu}-x_{(1)}^{\mu}\right)$ and the
average position $x_{(+)}^{\mu}=1/2\left(x_{(2)}^{\mu}+x_{(1)}^{\mu}\right)$.
It follows from our discussion above that under rigid translations,
with translation parameter $\xi^{\mu}=\left(\xi^{0},\bm{\xi}\right)$,
such that ${x'}_{(a)}^{\mu}=x_{(a)}^{\mu}+\delta_{\xi}x_{(a)}^{\mu}$,
the relative position does not change: $\delta_{\xi}x_{(-)}^{\mu}=0$.
Therefore the variation of the Lagrangian under infinitesimal translations
will be proportional to $\delta_{\xi}x_{(+)}^{\mu}$. Notice first
that, after rearranging the coordinates, the variation of the kinetic term is
\begin{equation}
\begin{split} & \delta_{\xi}{\cal L}_{1\oplus2}^{kin}=\delta_{\xi}x_{(+)}^{0}\left(\dot{p}_{(1)0}+\dot{p}_{(2)0}\right)\\
 & \ \ \ +\delta_{\xi}x_{(+)}^{j}\left[\dot{p}_{(1)j}+e^{-p_{(1)0}/\kappa}\dot{p}_{(2)j}-\frac{1}{\kappa}e^{-p_{(1)0}/\kappa}p_{(2)j}\dot{p}_{(1)0}+\frac{1}{\kappa}\left(p_{(1)j}+e^{-p_{(1)0}/\kappa}p_{(2)j}\right)\left(\dot{p}_{(1)0}+\dot{p}_{(2)0}\right)\right].
\end{split}
\label{eq:kin+}
\end{equation}
We want the coordinate variation to be $\delta_{\xi}x_{(+)}^{\mu}=-\xi^{\mu}+O\left(1/\kappa\right)$,
with coordinate-independent parameters $\xi^{\mu}$, in order to recover
the standard translation in the limit $1/\kappa\rightarrow0$. Let
us first consider terms in the variation proportional to $\xi^{0}$.
It is straightforward from\ (\ref{eq:kin+}), considering that the
term proportional to $\delta_{\xi^{0}}x_{(+)}^{0}$ is a total derivative,
that imposing
\begin{equation}
\delta_{\xi^{0}}x_{(+)}^{0}=-\xi^{0},\qquad\delta_{\xi^{0}}x_{(+)}^{j}=0,\label{eq:transx+0}
\end{equation}
it follows
\begin{equation}
\delta_{\xi^{0}}{\cal L}_{1\oplus2}^{kin}=-\xi^{0}\frac{d}{d\tau}\left(p_{(1)0}+p_{(2)0}\right),
\end{equation}
so that the zeroth component of the conserved total momentum is
\begin{equation}
p_{0}^{tot}=p_{(1)0}+p_{(2)0}.\label{eq:TotMom0}
\end{equation}

It is far less trivial to find the spacial component on the conserved
total momentum. From\ (\ref{eq:kin+}) one can notice that the terms
proportional to $\delta_{\bm{\xi}}x_{(+)}^{j}$ do not to add up to
a total derivative. However one can verify that by setting the variation
parametrized by $\bm{\xi}$ to be
\begin{equation}
\delta_{\bm{\xi}}x_{(+)}^{0}=\frac{1}{\kappa}\xi^{j}\left(p_{(1)j}+e^{-p_{(1)0}/\kappa}p_{(2)j}\right),\qquad\delta_{\bm{\xi}}x_{(+)}^{j}=-\xi^{j},\label{eq:transx+j}
\end{equation}
we get
\begin{equation}
\delta_{\bm{\xi}}{\cal L}_{1\oplus2}^{kin}=-\xi^{j}\frac{d}{d\tau}\left(p_{(1)j}+e^{-p_{(1)0}/\kappa}p_{(2)j}\right),
\end{equation}
so that the spacial component of the total momentum is
\begin{equation}
p_{j}^{tot}=p_{(1)j}+e^{-p_{(1)0}/\kappa}p_{(2)j}.\label{eq:TotMomj}
\end{equation}

One can check by~(\ref{eq:PB})-(\ref{eq:PBmixed}), that the total momentum
$p_{\mu}^{tot}$ (\ref{eq:TotMom0})\ (\ref{eq:TotMomj}), generates
the translations\ (\ref{eq:transx+0})\ (\ref{eq:transx+j}) by
Poisson brackets, its action on the single particle coordinates being
\begin{equation}
\begin{gathered}\delta_{\xi}x_{(a)}^{\mu}=-\left\{ \xi^{\nu}p_{\nu}^{tot},x_{(a)}^{\mu}\right\} ,\\
\delta_{\xi}x_{(1)}^{0}=\delta_{\xi}x_{(2)}^{0}=-\xi^{0}+\frac{1}{\kappa}\bm{\xi}\cdot\mathbf{p}^{tot},\qquad\delta_{\xi}x_{(1)}^{j}=\delta_{\xi}x_{(2)}^{j}=-\xi^{j}.
\end{gathered}
\label{eq:transChange}
\end{equation}
Both particles are translated by the same amount, consistently with
our assumptions of rigid translations. Notice now that the total momentum
generating rigid translations can be re-expressed as a deformed summation
law for the single particle momenta
\begin{equation}
\begin{gathered}
p_{\mu}^{tot}=\left(p_{(1)}\oplus p_{(2)}\right)_{\mu},\\
\text{with} ~~~~~ \left(p_{(1)}\oplus p_{(2)}\right)_{0} = p_{(1)0}+p_{(2)0},
~~~~  \left(p_{(1)}\oplus p_{(2)}\right)_{j} = p_{(1)j}+e^{-p_{(1)0}/\kappa}p_{(2)j} .
\end{gathered}
\label{eq:TotMomOplus}
\end{equation}
This deformed summation law, in turn, by means of~(\ref{eq:k-mink}), can be associated
with the product of the two particle group valued momenta\ (\ref{eq:TTR})
as
\begin{equation}\label{28}
{\cal P}^{tot}\left(p_{(1)}\oplus p_{(2)}\right)={\cal P}_{(1)}\left(p_{(1)}\right){\cal P}_{(2)}\left(p_{(2)}\right).
\end{equation}

\section{New action proposal for two-particle processes\label{sec:action}}

We can draw some considerations from the results of the previous section.
\begin{itemize}
\item The rigid translations are generated by the total momentum obtained
respectively by the group elements ${\cal P}^{tot}={\cal P}_{(1)}{\cal P}_{(2)}$.
This reflects the structure of the kinetic terms\ (\ref{eq:kineticTerm}),
where the momenta appear in the same combination. Notice indeed that
the kinetic term\ (\ref{eq:kineticTerm}) can be re-expressed as
\[
\left\langle \left[\frac{d}{d\tau}\left({\cal P}_{(1)}{\cal P}_{(2)}\right)\right]\left({\cal P}_{(1)}{\cal P}_{(2)}\right)^{-1}-\left(\frac{d}{d\tau}{\cal P}_{(2)}\right){\cal P}_{(2)}^{-1},x_{(1)}\right\rangle +\left\langle \left(\frac{d}{d\tau}{\cal P}_{(2)}\right){\cal P}_{(2)}^{-1},x_{(2)}\right\rangle .
\]

\item The change in the coordinates due to rigid translations can be rewritten
in terms of the total momentum as (\ref{eq:transChange}), i.e. the
particle coordinates translate rigidly in function of the total momentum.
\end{itemize}
Consider now a process involving two incoming particles $(1)$, $(2)$
and two outgoing particles $(3)$, $(4)$. Taking into account of
the above considerations and of the results reported in~\cite{AmelinoCamelia:2011nt}
(see also~\cite{Carmona:2011wc}), and later in~\cite{Amelino-Camelia:2014qaa},
where the analysis of translational invariant formulations of relative
locality frameworks is developed, and considering also that the two-particle
system evolution is parametrized by the external time $\tau$, we
propose the following action:
\begin{equation}
S=\int_{-\infty}^{\bar{\tau}}d\tau\left[{\cal L}_{1\oplus2}+\frac{d}{d\tau}\left(\zeta^{\mu}\left(p_{(1)}\oplus p_{(2)}\right)_{\mu}\right)\right]+\int_{\bar{\tau}}^{\infty}d\tau\left[{\cal L}_{3\oplus4}+\frac{d}{d\tau}\left(\zeta^{\mu}\left(p_{(3)}\oplus p_{(4)}\right)_{\mu}\right)\right].\label{eq:action2-2}
\end{equation}
Here the process is taken to happen at the time $\tau=\bar{\tau}$.
The interaction is described by the two boundary terms, where $\zeta^{\mu}$
is a function of $\tau$, $\zeta^{\mu}=\zeta^{\mu}\left(\tau\right)$.
The action differs from previous proposal since each composite particle system, the incoming and the outgoing, is described by its own action integral, characterized by the total momentum that generates the same rigid translations.
However one can show that for a single process, the boundary terms reduce to an interaction term of the kind reported\footnote{The two formulations differ only when one considers more than one process.
In that case in~\cite{AmelinoCamelia:2011nt} the translational invariance
is implemented in such a way that the change of the interaction coordinate
$\zeta^{\mu}$ under translation is the same for every process, $\delta\zeta^{\mu}=\xi^{\mu}$.
In~\cite{Carmona:2011wc} $\delta\zeta^{\mu}$ changes differently
for each process, the changes being functions of the momenta of all
the considered processes. The implications of this are discussed in~\cite{Amelino-Camelia:2014qaa}.} in~\cite{Carmona:2011wc} or in~\cite{AmelinoCamelia:2011nt}.
Indeed varying the action respect to $\zeta{}^{\mu}$, we get%
\footnote{We assume $\delta\zeta^{\mu}\left(-\infty\right)=\delta\zeta^{\mu}\left(\infty\right)=0$.%
}
\begin{equation}
\delta\zeta^{\mu}\left(\bar{\tau}\right)\left(\left(p_{(1)}\oplus p_{(2)}\right)_{\mu}-\left(p_{(3)}\oplus p_{(4)}\right)_{\mu}\right)=0
\end{equation}
i.e. we recover the constraint equation
\begin{equation}
p_{(in)\mu}^{tot}=\left(p_{(1)}\oplus p_{(2)}\right)_{\mu}=\left(p_{(3)}\oplus p_{(4)}\right)_{\mu}=p_{(out)\mu}^{tot}.
\label{eq:ConservationMomentaProcess}
\end{equation}

Even if the boundary terms of our model, for a single process, reduce to an interaction term of the kind described in~\cite{Carmona:2011wc} or~\cite{AmelinoCamelia:2011nt}, the structure of the kinetic terms is such that translations are substantially modified.
Variating the action respect to the particles momenta
$p_{(a)\mu}$, we get the boundary conditions for the worldlines endpoints
\begin{equation}
\begin{gathered}
x_{(a)}^{0}\left(\bar{\tau}\right)=-\zeta^{0}(\bar{\tau})+\frac{1}{\kappa}\bm{\zeta}(\bar{\tau})\cdot\mathbf{p}^{tot}_{(in)},\quad
x_{(a)}^{j}\left(\bar{\tau}\right)=-\zeta^{j}(\bar{\tau}),
\quad a=1,2, \\
x_{(a)}^{0}\left(\bar{\tau}\right)=-\zeta^{0}(\bar{\tau})+\frac{1}{\kappa}\bm{\zeta}(\bar{\tau})\cdot\mathbf{p}^{tot}_{(out)},\quad
x_{(a)}^{j}\left(\bar{\tau}\right)=-\zeta^{j}(\bar{\tau}),
\quad a=3,4.
\end{gathered}
\end{equation}
The Lagrange multiplier $\zeta^\mu$ can be interpreted as the coordinate of the vertex (cf.~\cite{AmelinoCamelia:2011bm}, \cite{Carmona:2011wc}, \cite{AmelinoCamelia:2011nt}).
The translations are implemented by variating these equations. Considering that momenta are unaffected by translations, and setting $\delta \zeta^\mu = \xi^\mu$, $\xi^\mu$ being the translation parameter, we obtain the relations
\begin{equation}
\begin{gathered}
\delta x_{(a)}^{0}\left(\bar{\tau}\right)=-\xi^0+\frac{1}{\kappa}\bm{\xi}\cdot\mathbf{p}^{tot}_{(in)},\quad
x_{(a)}^{j}\left(\bar{\tau}\right)=-\xi^{j},
\quad a=1,2, \\
\delta x_{(a)}^{0}\left(\bar{\tau}\right)=-\xi^{0}+\frac{1}{\kappa}\bm{\xi}\cdot\mathbf{p}^{tot}_{(out)},\quad
x_{(a)}^{j}\left(\bar{\tau}\right)=-\xi^{j},
\quad a=3,4.
\end{gathered}
\label{eq:translationProcess}
\end{equation}
Thus, from the last equations and the conservation equation (\ref{eq:ConservationMomentaProcess}), we get that the the coordinates of the endpoints of both incoming and outgoing particle worldlines translate rigidly by the same amount, i.e. proportionally to the conserved total momentum of the system. It is clear that if the interaction is local for the first observer, i.e. if the worldlines endpoints coincide for the first observer, they coincide also for a translated observer, so that the locality of the interaction is preserved under translations.

Comparing Eq.~(\ref{eq:translationProcess}) to (\ref{eq:transChange}), one can see that the change in the worldlines endpoints is exactly the one generated by the total momentum of the system by Poisson brackets. Extending this property to all the worldlines points, we can express the translation in terms of Poisson brackets as
\begin{equation}
\delta x_{(a)}^{\mu}=-\xi^{\nu}\left\{ p_{(in)\nu}^{tot},x_{(a)}^{\mu}\right\} ,\quad a=1,2\qquad \delta x_{(a)}^{\mu}=-\xi^{\nu}\left\{ p_{(out)\nu}^{tot},x_{(a)}^{\mu}\right\},\quad a=3,4.
\end{equation}

As it is well known there is, generally, some freedom in the form of the momentum conservation law, in the case of the nonlinear momentum composition. Indeed, any change in the ordering of momenta gives a new condition for components of the momenta conservation, while neither choice is better than other. In our case this ambiguity is reduced considerably. First of all the ordering of the kinetic term gives $p^{tot} = p_{(1)} \oplus p_{(2)}$ and not the other way around. Secondly, since our total momentum is a generator of rigid translation, the only consistent way to implement the momentum conservation is to demand that the translation before and after the vertex must be exactly the same, from which it follows that the initial total momentum must be equal to the final one, Eq.~(\ref{eq:ConservationMomentaProcess}).

The preservation of locality under translations represents the major difference with respect the previous formulations of relative locality. In the relative locality models described so far in the literature, a process appears to be local only for an observer whose coordinates origin coincide with the interaction point (i.e. an observer who is local to the process). Another observer, at rest relatively to the observer local to the process, but distant from the process, will describe the interaction as non-local.
The reason for that is that the change in the coordinates due to translation is different for each particle, proportional to the particle momentum. Then, for the second observer, distant from the vertex, the endpoints of the particles worldlines will not coincide, so that she will describe the distant process as non-local.

In our model, on the other hand, the translations are rigid. The change in the coordinates characterizing the endpoints of the particles worldlines will be the same for each particle, proportional to the conserved total momentum of the system. Then, if the endpoints coincide for the first observer, local to the process, i.e. they coincide in her frame's origin, they will coincide also for the second distant observer, at a point in her frame distant from her origin.
The locality of a particle process is preserved by translations.

Based on this, we could have concluded that in our formulation there is no relative locality for translations.
However, so far we have exhibited our result only for the case of a single vertex, while it has been shown~\cite{AmelinoCamelia:2011nt,Amelino-Camelia:2014qaa} that the most striking features of relative locality emerge when considering a combination of several processes, some of which causally connected\footnote{We refer here to the definition of causally connected processes adopted for instance in~\cite{AmelinoCamelia:2011nt,Amelino-Camelia:2014qaa}: the two processes are causally connected when an outgoing particle of one of the two processes is also an ingoing particle of the other.}.
While our proposal can be generalized to an arbitrary number of particles and interaction vertices, we postpone a detailed discussion of such a generalization and its physical properties to a forthcoming paper.

Let us still make a brief qualitative remark. Consider the detection (by a single detector) of two particles of different energy emitted simultaneously at a distant source. The event of emission and detection for each particle can be characterized by a suitable process. Imagine (schematically) the source to be composed of a large number of excited atoms, and imagine the two particles to be emitted by the simultaneous decay of two atoms. Obviously the processes of emission of the two particles are not causally connected (in the sense defined above).
Similarly, the processes of detection of the two particles, which can be imagined as the excitation by the particles of two atoms belonging to the detection apparatus, are not causally connected.
The process of emission and the process of detection associated with the same particle are instead causally connected: each detected particle is both the ingoing particle for its own detection process and the outgoing particle for its own emission process.
A similar setup for the relative locality framework has been discussed for instance in~\cite{AmelinoCamelia:2011nt,Amelino-Camelia:2014qaa}.

One can show (we will discuss this point in detail in a forthcoming paper) that in our model, the translation of each chain of causally connected processes is characterized by its total momentum, in the same way as the 2-particle system discussed in here. Then, individually, the processes remain local under translations, in the sense that, for all translated observers, the endpoints of the worldlines characterizing each process coincide in the interaction point, as shown for the 2-particle process discussed in this section.
However, since each chain of causally connected processes translates proportionally to its own total momentum, the two events of detection (as well as the two events of emission), which are not causally connected, and are characterized by different total momentum, translate by a different amount. This means that while the processes of detection (and emission) taken individually remain local, in the sense that all endpoints characterizing the process converge in the interaction point, they still present relative locality effects between themselves, when comparing the positions of the two events. An observer, distant from the simultaneous emission of the two particles (characterized by two causally unconnected processes), will describe in her coordinatization the two events to happen at different spacetime positions.
Thus we expect the relative locality features to reappear once considering physical situations involving several processes.

\section{Boosts and rotations}

We saw in the previous sections that our action for a composite particle system is fully characterized by its total momentum, which in turn implies that the translations, generated by the total momentum of the system, are rigid, so that the locality of a single process is preserved under translation. We want now to study if a similar property is fulfilled by the remaining spacetime transformations, boosts and rotations.

\subsection{Algebra of symmetries for the two-particle system}

The mass shell relation for each particle of the two-particle system is the $\kappa$-Poincar\'e mass Casimir (\ref{eq:casimir}). The single-particle boost and rotation charges/generators compatible with this choice of Casimir satisfy the $\kappa$-Poincaré algebra~\cite{Majid:1994cy}
\begin{equation}
\begin{gathered}\left\{ {\cal N}_{(a)j},p_{(a)0}\right\} =p_{(a)j},\qquad\left\{ {\cal N}_{(a)j},p_{(a)k}\right\} =\delta_{jk}\left(\frac{\kappa}{2}\left(1-e^{-2p_{(a)0}/\kappa}\right)+\frac{1}{2\kappa}\mathbf{p}_{(a)}^{2}\right)-\frac{1}{\kappa}p_{(a)j}p_{(a)k},\\
\left\{ R_{(a)j},p_{(a)0}\right\} =0,\qquad\left\{ R_{(a)j},p_{(a)k}\right\} =\epsilon_{jkl}p_{(a)l},\\
\left\{ R_{(a)j},R_{(a)k}\right\} =\epsilon_{jkl}R_{(a)l},\qquad\left\{ R_{(a)j},{\cal N}_{(a)k}\right\} =\epsilon_{jkl}{\cal N}_{(a)l},\qquad\left\{ {\cal N}_{(a)j},{\cal N}_{(a)k}\right\} =-\epsilon_{jkl}R_{(a)l},
\end{gathered}
\label{eq:algebra}
\end{equation}
and they are conserved charges since from (\ref{eq:Hamiltonian}) it follows
\begin{equation}
\dot{{\cal N}}_{(a)j}=\left\{ {\cal H},{\cal N}_{(a)j}\right\} =0,\qquad\dot{R}_{(a)j}=\left\{ {\cal H},R_{(a)j}\right\} =0.
\end{equation}
Using relations (\ref{eq:PB}), (\ref{eq:PBmixed}) and (\ref{eq:PBcoord}), one finds the representation of the single-particle boost and rotation generators in terms of phase space variables to be
\begin{equation}
\begin{gathered}{\cal N}_{(1)j}=-p_{(1)j}x_{(1)}^{0}-\left(\frac{\kappa}{2}\left(1-e^{-2p_{(1)0}/\kappa}\right)+\frac{1}{2\kappa}\mathbf{p}_{(1)}^{2}\right)x_{(1)}^{j},\\
{\cal N}_{(2)j}=-p_{(2)j}x_{(2)}^{0}-\left(\frac{\kappa}{2}\left(1-e^{-2p_{(2)0}/\kappa}\right)+\frac{1}{2\kappa}\mathbf{p}_{(2)}^{2}\right)\left(x_{(2)}^{j}-\left(1-e^{-p_{(1)0}/\kappa}\right)x_{(1)}^{j}\right)-\frac{1}{\kappa}p_{(2)j}\mathbf{p}_{(1)}\cdot\mathbf{x}_{(1)},
\end{gathered}
\label{eq:boosts}
\end{equation}
\begin{equation}
R_{(1)j}=R_{(1)j}=\epsilon_{jkl}p_{(1)k}x_{(1)}^{l},\qquad R_{(2)j}=\epsilon_{jkl}p_{(2)k}\left(x_{(2)}^{l}-\left(1-e^{-p_{(1)0}/\kappa}\right)x_{(1)}^{l}\right),\label{eq:rotations}
\end{equation}

To derive the expression for the total boost and rotation generators, we notice that the two particles system is seen by an observer with not sufficient ``resolution power'' as a single system carrying the momentum $p^{(tot)}_\mu$. It follows that the total momentum should transform, with respect to the symmetries generated by total boost ${\cal N}^{(tot)}_{j}$ and rotation $R^{(tot)}_{j}$, in exactly the same way as the momenta $p_{(a)\mu}$ transform under the symmetries generated by the single particle boost ${\cal N}_{(a)j}$ and rotation $R_{(a)j}$, Eq.~(\ref{eq:algebra}). This property, in turn, ensures the covariance of the energy-momentum conservation law~(\ref{eq:ConservationMomentaProcess}) under the action of total boost and rotation (cf.~\cite{AmelinoCamelia:2011yi}).
Thus we want the total boost and total rotation generators to satisfy the property
\begin{equation}
\left\{ {\cal N}_{j}^{tot},p_{\mu}^{tot}\right\} =\left\{ {\cal N}_{(a)j},p_{(a)\mu}\right\} \Big|_{\substack{p_{(a)}\rightarrow p^{tot} \\ {\cal N}_{(a)}\rightarrow{\cal N}^{tot}}},
\quad
\left\{ R_{j}^{tot},p_{\mu}^{tot}\right\} =\left\{ R_{(a)j},p_{(a)\mu}\right\} \Big|_{\substack{p_{(a)}\rightarrow p^{tot} \\ R_{(a)}\rightarrow R^{tot}}}.
\end{equation}
One finds that the total boost and rotation generators satisfying this relation have the expression
\begin{equation}
{\cal N}_{j}^{tot}={\cal N}_{(1)j}+e^{-p_{(1)0}/\kappa}{\cal N}_{(2)j}+\frac{1}{\kappa}\epsilon_{jkl}p_{(1)k}R_{(2)l},\qquad R_{j}^{tot}=R_{(1)j}+R_{(2)j},\label{eq:2-partBoost}
\end{equation}
which is exactly the expression that follows from the
 $\kappa$-Poincaré coproducts~\cite{Majid:1994cy}.
We see therefore the remarkable property that the expressions for the  generators of total momentum, boost and rotation deduced in our classical model on the basic physical premises, reproduce the coproduct structure of the $\kappa$-Poincaré Hopf algebra.
One can verify that the property for the total generators to transform in the same way as the single-particle generators, i.e. for the total generators to satisfy the same algebra of the single-particle ones, is satisfied for all the set of spacetime symmetries: calling ${\cal G}_{(a)\mu}$ the generic element of the set of single-particle charge/generators (momenta, boosts, and rotations), and ${\cal G}_{\mu}^{tot}$ their two-particle composite version, one can verify with the help of relations (\ref{eq:algebra}), that the following property is satisfied:
\begin{equation}
\left\{ {\cal G}_{\mu}^{tot},{\cal G}_{\nu}^{tot}\right\} =\left\{ {\cal G}_{(a)\mu},{\cal G}_{(a)\nu}\right\} \Big|_{{\cal G}_{(a)}\rightarrow{\cal G}^{tot}}.
\end{equation}

\subsection{Action of boosts and rotations on a 2-particle system}

We now turn to examine the behavior of coordinate changes under the action of the total boost and rotation generators defined in the previous subsection.
Under the action of a composite boost and rotation, the spacetime
coordinates change respectively as
\begin{equation}
{x_{(a)}'}^{\mu}=x_{(a)}^{\mu}+\delta_{\lambda}x_{(a)}^{\mu}=x_{(a)}^{\mu}-\bm{\lambda}\cdot\left\{ \bm{{\cal N}}^{tot},x_{(a)}^{\mu}\right\} ,\qquad{x_{(a)}'}^{\mu}=x_{(a)}^{\mu}+\delta_{\theta}x_{(a)}^{\mu}=x_{(a)}^{\mu}-\bm{\theta}\cdot\left\{ \mathbf{R}^{tot},x_{(a)}^{\mu}\right\} ,
\label{coordTransformation}
\end{equation}
$\lambda^{j}$ and $\theta^{j}$ being the boost and rotation parameters.
The system of particle transforms under the action of the composite
boost and rotation (\ref{eq:2-partBoost}), as
\begin{equation}
\begin{gathered}\left\{ {\cal N}_{j}^{tot},x_{(1)}^{0}\right\} =-x_{(1)}^{j}-\frac{1}{\kappa}{\cal N}_{j}^{tot},\qquad\left\{ {\cal N}_{j}^{tot},x_{(1)}^{k}\right\} =-\delta_{jk}x_{(1)}^{0}+\frac{1}{\kappa}\epsilon_{jkl}R_{l}^{tot},\\
\left\{ {\cal N}_{j}^{tot},x_{(2)}^{0}\right\} =-e^{-p_{(1)0}/\kappa}x_{(2)}^{j}-\left(1-e^{-p_{(1)0}/\kappa}\right)x_{(1)}^{j}-\frac{1}{\kappa}{\cal N}_{j}^{tot},\\
\begin{aligned}\left\{ {\cal N}_{j}^{tot},x_{(2)}^{k}\right\} = & -\delta_{jk}\left(e^{-p_{(1)0}/\kappa}x_{(2)}^{0}+\left(1-e^{-p_{(1)0}/\kappa}\right)x_{(1)}^{0}-\frac{1}{\kappa}\mathbf{p}_{(1)}\cdot\left(\mathbf{x}_{(2)}-\mathbf{x}_{(1)}\right)\right)\\
 & +\frac{1}{\kappa}\epsilon_{jkl}R_{l}^{tot}-\frac{1}{\kappa}p_{(1)j}\cdot\left(x_{(2)}^{k}-x_{(1)}^{k}\right),
\end{aligned}
\end{gathered}
\end{equation}
\begin{equation}
\left\{ R_{j}^{tot},x_{(1)}^{0}\right\} =\left\{ R_{j}^{tot},x_{(2)}^{0}\right\} =0,\qquad\left\{ R_{j}^{tot},x_{(1)}^{k}\right\} =\epsilon_{jkl}x_{(1)}^{l},\qquad\left\{ R_{j}^{tot},x_{(2)}^{k}\right\} =\epsilon_{jkl}x_{(2)}^{l}.
\end{equation}
One can verify that the action of boosts and rotations on the composite
system satisfies the property
\begin{equation}
\left(\left\{ {\cal N}^{tot},x_{(1)}^{\mu}\right\} -\left\{ {\cal N}^{tot},x_{(2)}^{\mu}\right\} \right)\Big|_{x_{(1)}=x_{(2)}}=0,\qquad\left(\left\{ R^{tot},x_{(1)}^{\mu}\right\} -\left\{ R^{tot},x_{(2)}^{\mu}\right\} \right)\Big|_{x_{(1)}=x_{(2)}}=0.\label{eq:rigidBoost}
\end{equation}
Notice that the interaction point, for a first observer that describes the interaction as local, is identified by the condition ($x_{(1)} = x_{(2)}$), i.e. the coinciding endpoint of the two particles worldlines.
Then, considering from Eq.~(\ref{coordTransformation}) that the change on the coordinates is proportional to their Poisson brackets with the total generators, Eq.~(\ref{eq:rigidBoost}) implies that if the interaction point is local for a first observer, it remains local also for a boosted or rotated observer:
\begin{equation}
\text{if} ~~ x_{(1)}=x_{(2)} ~~ \text{then} ~~ x'_{(1)}=x'_{(2)} .
\end{equation} 
Thus, as it was in the case of translations generated by the total momentum, the locality of a distant process is preserved also by the total boosts and rotations defined in this section.

\section{Summary and discussion}

Starting from a construction of a multi-particle Lagrangian inspired
by 2+1D gravity coupled to point particles, we
have proposed a new expression for the action of a two-particle process
with $\kappa$-Minkowski/$\kappa$-Poincaré deformed relativistic
symmetries, and associated deformed composition-law of momenta. The
latter is encoded in the product of group valued momenta, which in
turn reflects the properties of the coproduct of the deformed symmetry
algebra. The particle process is characterized by the conservation
of the total momentum of the system of particles incoming the process,
which equals the total momentum of the system of outgoing particles.
For each composite system the total momentum is the deformed sum of
the single-particle momenta, reflecting the properties of the non-commutative
spacetime symmetries, as required for a DSR description of a local
particle process~\cite{AmelinoCamelia:2000mn}.

The peculiarity of our action is that the two-particle coordinates
are ``braided'' together in the kinetic term, as a result of the
prescription coming from the 2+1D gravity analysis. Moreover we propose
a re-writing of the boundary term representing the vertex interaction
which takes into account of the properties of the momentum conservation
law in translational invariant formulations of ``principle of relative
locality'' theories (cf.~\cite{AmelinoCamelia:2011nt,Carmona:2011wc,Amelino-Camelia:2014qaa}).
In this way the action we obtain contains a single integral for each
composite particle system, with its own boundary term.

The main novelty of our action, respect to previous formulations~\cite{AmelinoCamelia:2011bm,Gubitosi:2013rna,Carmona:2011wc,AmelinoCamelia:2011nt},
is that the locality of a particle process is preserved by the whole
set of (deformed) relativistic transformations. One could conclude
that relativity of locality is not present in our model. However, we expect relative locality effects to emerge again when one compares causally unconnected chains of processes as explained in Sec.~\ref{sec:action}.

\section*{Acknowledgment}

This work was supported by funds provided by the National Science
Center under the agreement 2011/02/A/ST2/00294, and for JKG also by funds provided by the National Science Center under the agreement 2014/13/B/ST2/04043.


\end{document}